\documentclass[aps,prb,twocolumn,superscriptaddress,showpacs]{revtex4}
\usepackage{graphicx}

\begin{document}
\bibliographystyle{apsrev}

\title{Spin waves throughout the Brillouin zone and magnetic
exchange coupling in ferromagnetic metallic manganites
La$_{1-x}$Ca$_{x}$MnO$_3$ ($x=0.25,0.30$)}

\author{F. Ye}
\email{yef1@ornl.gov}
\affiliation{Neutron Scattering Science Division,
Oak Ridge National Laboratory, Oak Ridge, Tennessee 37831-6393,
USA}
\author{Pengcheng Dai}
\affiliation{Department of Physics and Astronomy,
The University of Tennessee, Knoxville, Tennessee 37996-1200, USA}
\affiliation{Neutron Scattering Science Division,
Oak Ridge National Laboratory, Oak Ridge, Tennessee 37831-6393,
USA}

\author{J. A. Fernandez-Baca}
\affiliation{Neutron Scattering Science Division,
Oak Ridge National Laboratory, Oak Ridge, Tennessee 37831-6393,
USA}
\affiliation{Department of Physics and Astronomy,
The University of Tennessee, Knoxville, Tennessee 37996-1200, USA}

\author{D. T. Adroja}
\affiliation{ISIS facility, Rutherford Appleton Laboratory,
Chilton, Didcot OX11 0QX, UK}

\author{T. G. Perring}
\affiliation{ISIS facility, Rutherford Appleton Laboratory,
Chilton, Didcot OX11 0QX, UK}

\author{Y. Tomioka}
\affiliation{Correlated Electron Research Center (CERC),
Tsukuba 305-0046, Japan}

\author{Y. Tokura}
\affiliation{Correlated Electron Research Center (CERC),
Tsukuba 305-0046, Japan}
\affiliation{Department of Applied Physics, University of Tokyo,
Tokyo 113-8656, Japan}

\date{\today}

\begin{abstract}
Using time-of-flight and triple-axis inelastic neutron spectroscopy,
we determine spin wave excitations throughout the Brillouin zone for
ferromagnetic manganites La$_{1-x}$Ca$_x$MnO$_3$ ($x=0.25,0.3$) in
their low temperature metallic states. While spin wave excitations
in the long wavelength limit (spin stiffness $D$) have similar
values for both compounds, the excitations near the Brillouin zone
boundary of La$_{0.7}$Ca$_{0.3}$MnO$_3$ are considerable softened in
all symmetry directions compared to that of
La$_{0.75}$Ca$_{0.25}$MnO$_3$. A Heisenberg model with the nearest
neighbor and the fourth neighbor exchange interactions can describe
the overall dispersion curves fairly well. We compare the data with
various theoretical models describing the spin excitations of
ferromagnetic manganites.
\end{abstract}


\maketitle
\narrowtext
\section{Introduction}

The experimental investigation of spin dynamical properties in doped
manganese perovskite $A_{1-x}B_{x}$MnO$_3$, where $A$ is the
trivalent ion ($\rm La^{3+}, Pr^{3+}, Nd^{3+}$, etc) and $B$ is the
divalent ion ($\rm Ca^{2+}$ or $\rm Sr^{2+}$), is essential to the
understanding of spin-spin interactions in these materials. At
hole-doping level $x\approx 0.30$, these so-called colossal
magnetoresistance (CMR) compounds exhibit an unusually large change
in electrical resistance in response to a magnetic field and changes
from a paramagnetic to a ferromagnetic
state.\cite{salamon,tokurabook00,dagotto01} The Mn $3d$ levels in
the $\rm Mn^{3+}/Mn^{4+}$ mixed valent system, split by the oxygen
octahedral crystal field to a lower energy $t_{2g}$ triplet and a
higher energy $e_g$ doublet, are filled according to the Hund's rule
such that all spins are aligned on a given site by a large
intra-atomic exchange coupling $J_H$. The basic microscopic
mechanism responsible for the CMR effect is the double-exchange (DE)
interaction,\cite{zener51} where ferromagnetism and electrical
conductivity arise from hopping of the itinerant $e_g$ electrons
with kinetic energy $t$ from trivalent $\rm Mn^{3+}$ to tetravalent
$\rm Mn^{4+}$ sites. In its simplest form, the Hamiltonian of a DE
model can be described as a single band of itinerant $e_g$ electrons
interacting with localized core spins by the Hund's rule exchange
$J_H$. Since $J_H$ is much larger than $t$, the kinetic energy of
itinerant $e_g$ electrons is minimal when all electron spins are
parallel, {\it i.e.}, the ground state is a metallic ferromagnet.

Although DE interaction is believed to be responsible for the
ferromagnetism and electron conductivity in CMR compounds, whether
the magnetic excitations of such a model can be discussed in terms
of an equivalent ferromagnetic Heisenberg model is still not clear.
In the strong coupling limit ($t/J_H \rightarrow 0$), Furukawa
\cite{furukawa96} has shown that the DE model can be mapped onto the
Heisenberg Hamiltonian with only the nearest neighbor (NN) exchange
coupling. In this scenario, the magnitude of the exchange coupling
$J$ associated with ferromagnetic spin waves should scale with the
Curie temperature $T_c$, kinetic energy $t$ and doping $x$.
\cite{martin96,perring96,endoh97,motome03} Experimentally, the
initial measurements on $\rm La_{0.7}Pb_{0.3}MnO_3$ suggest that a
simple NN Heisenberg model is sufficient to account for the entire
spin wave dispersion relation\cite{perring96} and the exchange
coupling obtained from such model also yields, to within 15\%, the
correct $T_c$ of the compound. However, later experimental
measurements indicate that spin wave excitations of most
$A_{1-x}B_x$MnO$_3$ manganites with $x\approx 0.3$ renormalize near
the zone boundary (ZB) with large softening and
damping.\cite{hwang98,doloc98,dai00phonon,chatterji02,moussa03}
Furthermore, the spin wave stiffness constant $D$ and the NN
magnetic exchange coupling $J$ are weakly dependent on $T_c$ and
doping level $x$ in metallic ferromagnetic (FM)
manganites.\cite{jaime98,dai01,ye06}

It is now well established that a Heisenberg model with NN exchange
coupling is insufficient to describe the dispersion relation of CMR
manganites, several possible microscopic mechanisms have been
proposed to address the unusual features of spin wave excitations near
the ZB. First, realistic calculations based on the DE mechanism
with consideration of the finite kinetic energy $t$ and the effect
of on-site Coulomb repulsion show that spin waves in the DE model do
not map to the Heisenberg Hamiltonian with simple NN exchange
coupling.\cite{golosov05} This model, however, predicts a doping
dependence on $D$ which is not observed experimentally. 
Second, the observed ZB spin wave softening may be due to the
conduction electron band filling effect, where the existence of
long-range magnetic interactions leads to the softening at the
ZB.\cite{solovyev99} On the other hand, whether this approach is
capable of explaining the observed spin wave broadening remains
unclear. Third, large magnon-phonon interactions may give rise to the
remarkable ZB softening along specific directions.\cite{dai00phonon}
Fourth, the deviation of short wave length magnons from the canonical
Heisenberg form might originate from the scattering of magnons by
collective quantum orbital fluctuations, which could either be
planar ($x^2-y^2$)-type orbitals associated with the A-type
antiferromagnetic (AF) ordering\cite{khaliullin00,krivenko04} or
rodlike ($3z^2-r^2)$ orbital correlations related to C-type AF
ordering.\cite{endoh05} Depending on the actual orbital shape, the
coupling between charge and orbital-lattice will give rise to
distinct doping dependence of the softening/broadening of the
magnetic spectrum. Fifth, the randomness created by the substitution
of the divalent ions for the trivalent ions in $A_{1-x}B_x$MnO$_3$
might be responsible for the anomalous spin wave
softening.\cite{motome02, motome05} Finally, the overlap between the
magnon excitations and Stoner continuum in the metallic
$A_{1-x}B_x$MnO$_3$ would cause softening and broadening of the
magnon branch near ZB.\cite{wang98,kaplan01} Although this
single-band DE model with intermediate coupling can explain the
softening/broadening in the low-$T_c$ compounds, remarkable
similarities in systems with widely different $T_c$'s indicate it is
inadequate as the bandwidth of Stoner Continuum is directly related
to the $T_c$'s.

Given that there are so many possible models to explain the ZB
magnon softening, it is imperative to carry out systematic spin wave
measurements and compare the results with predictions of various
models. In a recent Letter \cite{ye06}, we made such comparison for
spin waves along the $[\xi,0,0]$ direction and found that none of
prevailing models can account for the data. In this article, we
expand our previous work and describe a systematic investigation of
spin wave excitations of the CMR manganites $\rm
La_{0.75}Ca_{0.25}MnO_3$ (LCMO25) and $\rm La_{0.70}Ca_{0.30}MnO_3$
(LCMO30). Using reactor based and time-of-flight inelastic neutron
scattering (INS) techniques, we were able to map out the low
temperature ferromagnetic spin wave excitations of LCMO25 and LCMO30
throughout the Brillouin zone in all symmetry directions. In the
long wavelength limit, spin wave stiffness of LCMO25 and LCMO30 are
$147\pm3$ and $\rm 169\pm2~meV \AA^2$ respectively, consistent with
previous results \cite{dai01,ye06}. At large wavevectors, we find
that the dispersion relations of LCMO30 are considerable more
renormalized (softened) in all major symmetry directions compared to
those of LCMO25. The softening is well described by the introduction
of the $\rm 4^{th}$ NN ferromagnetic exchange coupling $J_4$
[Fig.~1(b)], the ratio of $J_4/J_1$ is about 19.5\% in LCMO30 and
6.5\% in LCMO25. In section II, we describe experimental details.
Section III gives the data analysis and comparison with previous
work. The conclusions are summarized in Section IV.

\section{Experimental Details}

We grew single crystals of LCMO25 and LCMO30 using the traveling
solvent floating zone technique. The Curie temperatures of LCMO25
($T_c = 190\pm1$~K) and LCMO30 ($T_c = 238\pm1$~K) are determined
from the elastic neutron diffraction on the (100) and (110) magnetic
Bragg peaks.\cite{dai00polaron} LCMO25 has a nominal hole doping
level of $x =0.25$, just above the metal-insulator transition
concentration ($x =0.22$). LCMO30 has a doping level close to
optimal doping with highest $T_c$. Our INS experiments were
performed on the HET chopper spectrometer at the ISIS spallation
neutron source, Rutherford- Appleton Laboratory, and on the HB1/HB3
triple-axis spectrometers at the High-Flux-Isotope Reactor (HFIR),
Oak Ridge National Laboratory. The momentum transfer wavevectors $q
=(q_x,q_y,q_z)$ are in units of $\AA^{-1}$ at positions
$(H,K,L)=(q_xa/2\pi,q_yb/2\pi,q_zc/2\pi)$ in reciprocal lattice
units (rlu), where $a\approx b\approx c\approx 3.87$ \AA\ and $3.86$
\AA\ are the lattice parameters of the pseudocubic unit cells of
LCMO25 and LCMO30, respectively [Fig.~1(a)]. The samples were
aligned in the $(H,H,L)$ zone in both the ISIS and HFIR
experiments. For the ISIS measurements, we use the HET
direct-geometry chopper spectrometer which has the $^3$He filled
detector tubes covering the scattering angles from 9-29$^\circ$ (PSD
detectors covers from 2.5-7.9$^\circ$). For the HFIR experiment, we
use pyrolytic graphite as monochromator, analyzer, and filters, and
the final neutron energy was fixed at $E_f = 13.5$ or 14.7~meV.

\begin{figure}[ht!]
\includegraphics[width=3.2in]{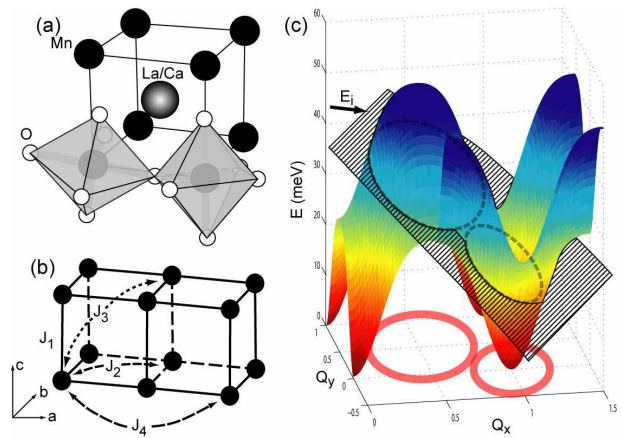}
\caption{\label{fig:diagram}
(Color online) (a) Crystal structure of $\rm La_{1-x}Ca_{x}$MnO$_3$
with distorted oxygen octahedra surrounding Mn ions. (b) Magnetic
exchange interactions up to $\rm 4^{th}$ order between adjacent Mn
ions. (c) Illustration of time-of-flight experiment. The
energy-wavevector ($E$-$q$) region probed by the experiment
intersects the spin wave dispersion surface, giving rise to spin
wave rings projected on the scattering plane. For smaller energy
transfer $E$, the ring centers at the zone center. For larger $E$,
the ring centers at the ZB.
}
\end{figure}

For the HET measurements, the magnetic scattering intensities of the
raw data were normalized to a vanadium standard. The scattering function
\begin{equation}
S({\bf{q}},\omega)=\frac{|k_i|}{|k_f|}\frac{d^2\sigma}{d\Omega d\omega},
\end{equation}
where $k_i$ and $k_f$ are the initial and final neutron wavevectors,
respectively, the solid angle of scattering is $\Omega$, and the
energy transfer is $\hbar\omega$. We used the program {\sc
mslice}\cite{mslice} to visualize the ${\bf q}$-$\omega$ data sets
and to prepare the one-dimensional cuts along the high-symmetry spin
wave directions, needed for further analysis using the program {\sc
tobyfit}\cite{tobyfit}. 

\section{Modeling and analysis}
The neutron scattering cross-section per formula unit
(f.u.) for spin wave excitations is
\begin{equation}
\frac{d^2\sigma}{d^2\Omega d\omega}=(\gamma
r_0)^2\frac{|k_f|}{|k_i|}|F({\bf{q}})|^2\frac{1}{\pi g^2
\mu^2_B}\frac{1}{1-e^{-\beta \hbar \omega}} \chi''({\bf{q}},\omega),
\end{equation}
where $(\gamma r_0)^2=0.2906~barn$, $|F({\bf{q}})|^2$ is the
magnetic form factor, $g$ is the Lande factor ($\approx 2$), $
[n(\omega)+1]=1/[1-\exp(-\hbar \omega/k_BT)]$ is the detailed
balance factor and $\mu_B$ is the Bohr magneton.
$\chi''({\bf{q}},\omega)$ is the imaginary part of the generalized
spin susceptibility which depends on the wavevector $q$ and energy
transfer $\hbar \omega$. In the damped simple harmonic oscillator
(DSHO) approximation, the normalized dynamical susceptibility
$\chi''({\bf{q}},\omega)$ can be written as
\begin{equation}
\chi''({\bf{q}},\omega)=\frac{4\gamma \omega \omega_0}
{\pi[(\omega^2-\omega^2_0)^2+4(\gamma\omega)^2]},
\label{eqn:chi}
\end{equation}
where $\gamma$ characterizes the damping of the magnetic spins,
$\omega_0$ is directly associated with the spin wave dispersion
relation. In the light damping limit, the intrinsic peak height $A$
and width $\Gamma$ of the spin wave excitation profiles are
determined by the damping $\gamma$ and become
\begin{equation}
A \propto 1/(2\pi \gamma), \Gamma \propto 2 \gamma.
\label{eqn:damping}
\end{equation}

The Hamiltonian for a Heisenberg ferromagnet is
\begin{equation}
H=-\frac{1}{2}\sum_{i,k}J_k {\bf S}_i \cdot {\bf S}_{i+k},
\label{eqn:hamiltonian}
\end{equation}
where ${\bf S}_i$ denotes the magnetic moment at site $i$, and $J_k$
indicates the magnetic exchange coupling between neighboring sites.
In an early study, a NN coupling $J_1$ has been successfully
employed to describe the entire spin wave dispersion relation
$E({\bf{q}})$ for high-$T_c$ manganites.\cite{perring96} Subsequent
measurements have shown the presence of ZB spin wave softening for
all other manganites with dispersions being reproduced well by a
Heisenberg Hamiltonian with higher order interactions.
\cite{hwang98,endoh05,ye06}
Recently, we found that the
introduction of the $\rm 4^{th}$ NN interactions $J_4$ gives
satisfactory description of ZB softening of the spin wave
dispersions for a wide range of doped manganites along the
$[\xi,0,0]$ direction.\cite{ye06}
We show here that such model also gives reasonable description
of spin waves in all other symmetry directions.

\subsection{Results on $\rm La_{0.75}Ca_{0.25}MnO_3$}

LCMO25 undergoes a ferromagnetic phase transition and becomes
metallic below $T_c=190$~K.\cite{dai00polaron} To determine the
dispersion of LCMO25 at large momentum transfers, we measured its
ferromagnetic spin waves with incident beam neutron energies ($E_i$)
of 32, 50, 75, 100, 125, 150, 175, and 185~meV at $T=8.5$~K (0.045
$T_c$) on HET. The sample was oriented such that either the [1,1,0]
or [0,0,1] axis of the crystal is along the incident beam
directions. For the neutron beam along the [1,1,0] axis of the
crystal, we could get the dispersion relations along the $[\xi,0,0]$
or $[\xi,\xi,0]$ directions. For the neutron beam along the [0,0,1]
axis, the dispersion along the $[\xi,\xi,\xi]$ direction can be
obtained.

\begin{figure}[ht!]
\includegraphics[width=3.2in]{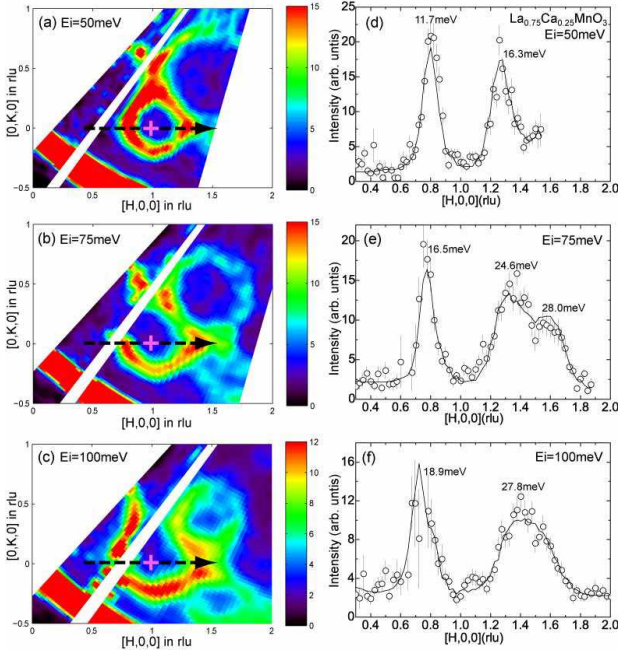}
\caption{\label{fig:lcmo25h00}
(Color online) Spin wave excitations of LCMO25 in the $[\xi,0,0]$
direction. Panels (a-c) illustrate the 2D contour plots in the
reciprocal space with incident energy of 50, 75, and 100~meV. The
cross symbols indicate the FM zone center of (1,0,0). Panels (d-f)
depict the corresponding scan profiles along the $[\xi,0,0]$ direction
shown by the arrows in the left panels. The typical cut width is 0.1
rlu. Solid lines are least square fits using the spin wave model
described in the text. The energies associated with excitation peaks are
labeled.
}
\end{figure}

Fig.~\ref{fig:lcmo25h00} summarizes the spin wave excitations for
incident neutron energies of 50, 75, and 100 meV. Panels (a-c) show
the two-dimensional color-coded contour plots of excitations in
reciprocal space. Two rings of scattering are observed in  these
panels. The first and strongest of these two is centered at the
(1,0,0) and corresponds to the intersection of the $E$-$q$ region
probed by the experiment, and the spin wave dispersion surface near
the zone center.  The second of these rings, centered at
(1.5,0.5,0), corresponds to such intersections near the ZB because
of a larger energy transfer $E$ [Fig.~1(c)].  We cut the $E$-$q$
data along the $[\xi,0,0]$ direction, as shown in panels (d-f). The
cut clearly shows two distinct peaks located at 11.7 and 16.3~meV
for $E_i=50$~meV [Fig.~2(d)]. These peaks gradually disperse outward
with increasing incident beam energy [Figs.~2(e),(f)].  As a
function of increasing energy and approaching the ZB wavevector, the
spin waves become broader in width and weaker in intensity
[Fig.~\ref{fig:lcmo25h00}(f)]. These results are consistent with our
earlier measurements.\cite{ye06}

\begin{figure}[ht!]
\includegraphics[width=3.2in]{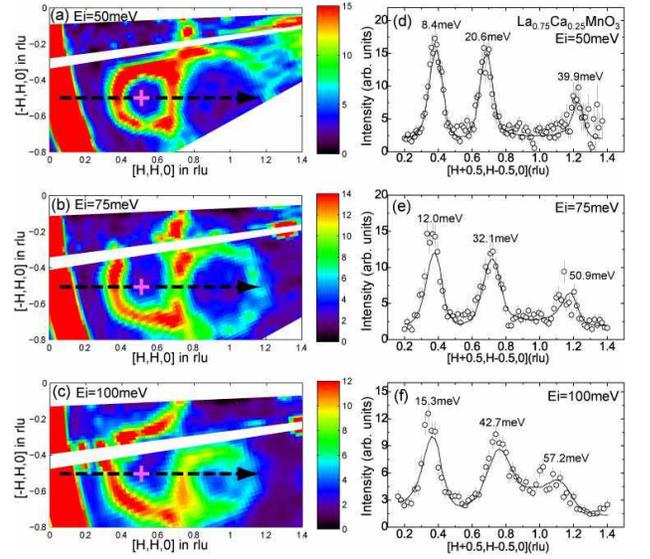}
\caption{\label{fig:lcmo25hh0}
(Color online) Spin wave excitations of LCMO25 in the $[\xi,\xi,0]$
direction with incident energies of 50, 75, and 100~meV. The viewing
axis of panels (a-c) are rotated $45^\circ$ with respect to those in
Fig.~\ref{fig:lcmo25h00} such that the scan profiles along
$[\xi,\xi,0]$ can be obtained. The cross symbols indicate the FM
zone center of (1,0,0). The dashed lines depict the cutting
direction.
}
\end{figure}

To obtain spin wave excitations along the $[\xi,\xi,0]$ direction,
we change the orthogonal viewing axes of the two-dimensional spin
wave spectra and cut images along the $[\xi,\xi,0]$ direction.
Fig.~\ref{fig:lcmo25hh0} summarizes the outcome of these cuts, it
is clear that the spin wave peaks become broader and weaker near the ZB
[Fig.~\ref{fig:lcmo25hh0}(c),(f)].

We systematically cut the data along all high symmetry directions
for various incident beam energies. To obtain reliable exchange
coupling constant $J_i$'s and determine their directional
dependence, we analyze data along one symmetry direction at a time
with as many cuts as possible. The data are fit simultaneously using
the dynamic susceptibility $\chi''(\omega,{\bf{q}})$ described in
Eqn.~\ref{eqn:chi} with spin wave dispersion relation as
\begin{equation}
\hbar \omega_0({\bf{q}})=\Delta+2S[J(0)-J({\bf{q}})],
\label{eqn:dispersion}
\end{equation}
where $\Delta$ is the anisotropic gap and
\begin{equation}
J({\bf{q}})=\sum_i J_k e^{i{\bf{q}}\cdot ({\bf{R}}_i-{\bf{R}}_j)}.
\label{eqn:exchange}
\end{equation}

Using the {\sc tobyfit} non-linear least-square analysis program, we
fit all the cuts by adjusting the peak amplitude $A$, damping term
$\Gamma$ and magnetic exchange coupling constant $J_i$'s. The
time-of-flight measurements do not provide information at small
momentum transfers, but the data obtained using triple-axis
spectroscopy showed negligible anisotropic spin gap near the zone
center.\cite{hwang98,doloc98,dai00phonon,chatterji02,moussa03} We
therefore fixed $\Delta=0$ in Eqn.~\ref{eqn:dispersion} during the
analysis.  Following the results of our previous paper, we force $J_2$ and
$J_3$ to be zero during the course of analysis, allowing only $J_1$
and $J_4$ to vary.\cite{ye06} Table~\ref{tab:lcmo25} shows the
fitting results in LCMO25 along three major symmetry directions of
$[\xi,0,0]$, $[\xi,\xi,0]$ and $[\xi,\xi,\xi]$. We note that the
value of exchange constants varies in a narrow range for fitted
results along different directions. For example, the NN exchange
coupling $J_1$ changes from 7.6 to 8.4~meV and $J_4$ varies from
0.26 to 0.58~meV.  Using $D=\bigtriangledown^2_{\bf{q}}
\omega_0({\bf{q}})|_{{\bf{q}}=0}$, where $\omega_0({\bf{q}})$ is the
dispersion relation, we can calculate the spin wave stiffness
constant $D$ and found that $D \approx 145$~meV \AA$^2$, consistent with
the value obtained from low-$q$ inelastic scattering
measurement.\cite{ye06} Since the fitting results along the three
high symmetry directions are consistent, we included excitation data
along all directions in the {\sc tobyfit}. As shown in
Table~\ref{tab:lcmo25}, the global analysis with as many as 38 data
sets along all directions gives $J_1=7.83\pm0.06$ and
$J_4=0.51\pm0.03$~meV, which yields the ratio of
$J_4/J_1=0.065\pm0.004$ and stiffness constant $D=147.2\pm2.6$~$\rm
meV \AA^2$.

\begin{table*}[ht!]
\begin{tabular}{llllccr}
\hline
\hline
Direction & $J_1$(meV) & $J_4$(meV)& $J_4/J_1$ & $D$($\rm meV \AA^2)$ & No.~of data sets &$\chi^2$\\
\hline
$[\xi,0,0]$    &7.56(7) &0.58(4) & 0.076(6) & $147.1\pm3.5$ & 11 & 1.45\\
$[\xi,\xi,0]$      &7.99(9) &0.44(4) &0.055(6) & $145.6\pm3.9$ & 13 & 1.81 \\
$[\xi,\xi,\xi]$    &8.39(14)&0.26(8) &0.032(10) & $140.7\pm6.6$ & 14 & 2.09 \\
$[\xi,0,0]+[\xi,\xi,0]+[\xi,\xi,\xi]$&7.83(6)&0.51(3) &0.065(4) & $147.2\pm2.6$ & 38 & 1.90 \\
$[\xi,0,0]+[\xi,\xi,0]+[\xi,\xi,\xi]$&8.75(3)&0(Fixed)&0   & $130.3\pm0.5$ & 38 & 2.32 \\
\hline
\hline
\end{tabular}
\caption{\label{tab:lcmo25}
Fitting parameters of the exchange coupling constant $J_1$, $J_4$, the
ratio of $J_4/J_1$ and spin wave stiffness constant $D$ in LCMO25.
Each data set is one dimensional scan profile as shown in
Fig.~\ref{fig:lcmo25h00}(d-f) and consists of one or more magnetic
excitation peaks. The results with only the NN interaction $J_1$ are
listed in the last row for comparison.}
\end{table*}

\begin{figure*}
\includegraphics[width=5.3in]{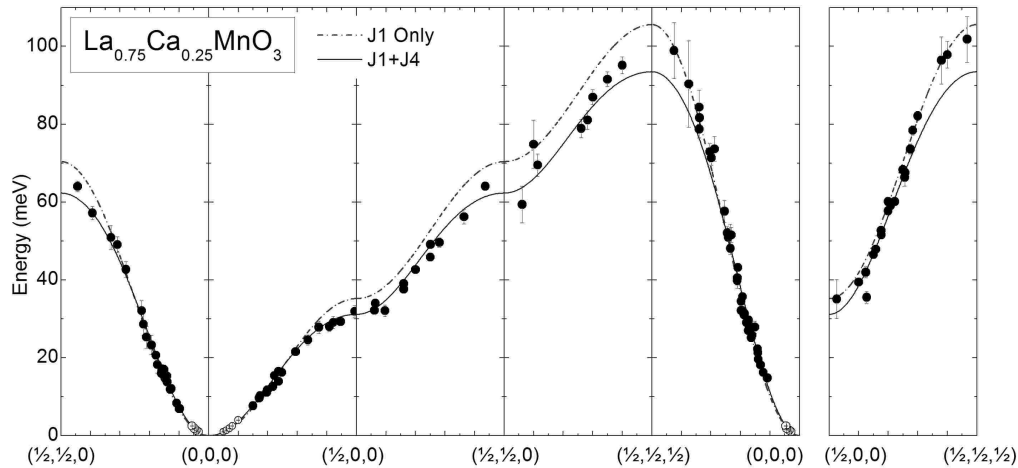}
\caption{\label{fig:lcmo25dispersion}
(Color online) The spin wave dispersion curves for LCMO25. The
solid points are data collected at ISIS and the open points at low
energies are collected at HFIR. The dash (red) curves are the fits
using only the NN interaction $J_1$. The solid (blue) curves are the
fits combining the NN interaction $J_1$ and the $\rm 4^{th}$ NN
interaction $J_4$ (see text).
}
\end{figure*}

{\sc tobyfit} gives only the fitting parameters $J_i$'s, $A$ and
$\Gamma$ for the spin wave dispersion relations. To actually
construct a plot of $E$ versus $q$ of LCMO25, we derive the energy
at an individual wavevector $E(q)$ by analyzing each single
one-dimensional cuts as shown in
Figs.~\ref{fig:lcmo25h00}(d-f).\cite{fitdetail}  As shown in
Fig.~\ref{fig:lcmo25dispersion}, the experimental data collapse
nicely onto the solid line generated using Eqns.~\ref{eqn:dispersion}
and \ref{eqn:exchange} with inclusion of the NN interaction $J_1$
and $\rm 4^{th}$ NN interaction $J_4$.  Finally, we show that a fit
to a purely NN exchange coupling $J_1$ is not adequate. The
dispersion relations based only on $J_1$ are plotted as the dashed
lines, although these curves describe reasonable well the low-$q$
data, they clearly deviate the data points at the ZB with higher
energies. For example, the actual ZB energies along $[\xi,0,0]$ and
$[\xi,\xi,0]$ directions are 31.0 and 62.5~meV, respectively. They
are lowered by about 4 and 8~meV from the prediction of the NN
Heisenberg Hamiltonian. It is noted that while the description using
only the NN exchange coupling is not sufficient to characterize the
entire dispersion relations in LCMO25, this exchange constant $J_1$
is similar to the value established in the INS study of $\rm
La_{0.7}Pb_{0.3}MnO_3$.\cite{perring96}

\subsection{Results on $\rm La_{0.7}Ca_{0.3}MnO_3$}

\begin{figure}[ht!]
\includegraphics[width=3.2in]{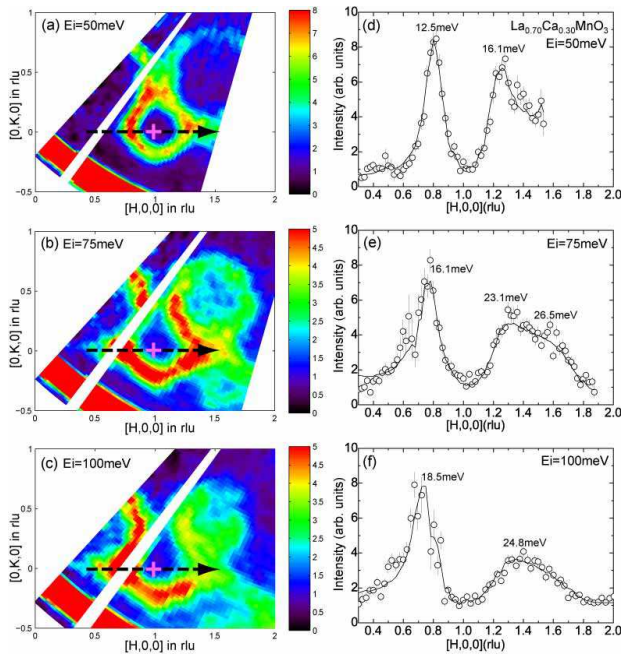}
\caption{\label{fig:lcmo30h00}
(Color online) Spin wave excitations of LCMO30 in the [H,0,0]
direction. Panels (a-c) illustrate 2D contour plots with
incident energies of 50, 75, and 100~meV. Solid lines are least square
fits. See Fig.~\ref{fig:lcmo25h00} for additional information.
}
\end{figure}

\begin{figure}[ht!]
\includegraphics[width=3.2in]{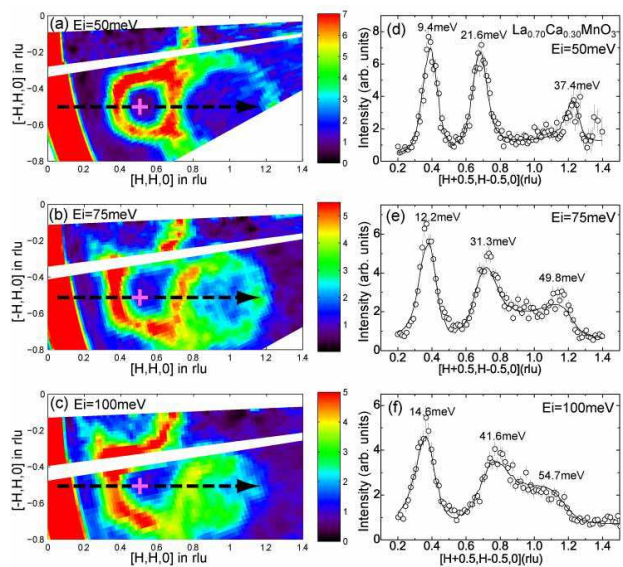}
\caption{\label{fig:lcmo30hh0}
(Color online) Spin wave excitations of LCMO30 in the [H,H,0]
direction with incident energies of 50, 75, and 100~meV.
See Fig.~\ref{fig:lcmo25h00} for additional information.
}
\end{figure}

To determine the evolution of spin wave excitations of ferromagnetic
CMR compounds, we also measured LCMO30 using HET. For these
measurements, we used incident beam energies of 50, 75, 100, 125,
150 and 185 meV. Fig.~\ref{fig:lcmo30h00} and
Fig.~\ref{fig:lcmo30hh0} show raw data for LCMO30 with the same
incident energies as those of LCMO25 in Figures~\ref{fig:lcmo25h00}
and \ref{fig:lcmo25hh0}. It is clear that LCMO30 has a softened
dispersion near the ZB. As shown in Fig.~\ref{fig:lcmo30h00}(f), the
peak height at $q=[1.4,0,0]$ is only half of that at $q=[0.7,0,0]$,
compared with 80\% for that in LCMO25 [Fig.~2(f)]. From Eqn.~4, the
decrease in peak height at a similar wavevector indicates a larger
damping term $\gamma$. In addition, the spin wave ring away from
(1.5,0.5,0) [Fig.~\ref{fig:lcmo30hh0}(c),(f)] collapses much faster
than in LCMO25, as the incident beam energy increases to 100~meV.
Table~\ref{tab:lcmo30} lists the quantitative fitting parameters
using {\sc tobyfit}. Much reduced $J_1$ and considerable enhanced
$J_4$ are observed along different high symmetry directions for
LCMO30. Similar to LCMO25, the softening near the boundary shows
little symmetry directional dependence. For the global fit, we
obtained $J_1=6.36\pm0.03$ and $J_4=1.24\pm0.02$~meV. The ratio of
$J_4/J_1$ reaches 20\%, a value much larger than that in LCMO25. The
stiffness constant is calculated to be $D=\rm 169\pm2~meV \AA^2$,
agrees well with the early triple-axis scattering
result.\cite{lynn96,dai01}

\begin{figure*}
\includegraphics[width=5.3in]{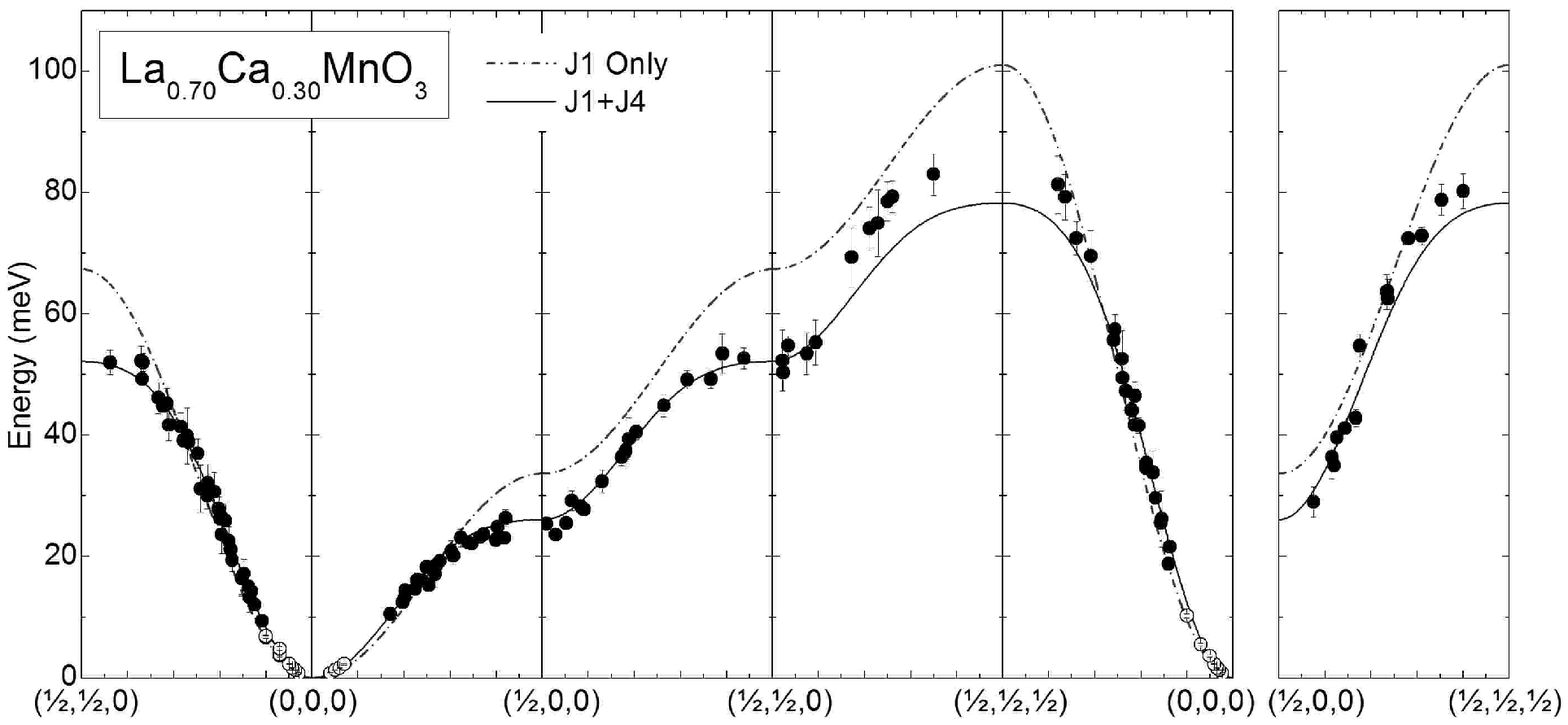}
\caption{\label{fig:lcmo30dispersion}
(Color online) The spin wave dispersion curves for LCMO30. The dash
(red) curves are the fits using only NN interaction $J_1$. The solid
(blue) curves are the fits using $J_1$ plus the $\rm 4^{th}$ NN
interaction $J_4$.
}
\end{figure*}

The dispersion curves covering the entire Brillouin Zone of LCMO30
are shown in Fig.~\ref{fig:lcmo30dispersion} with the solid lines
representing the fits using $J_1$ and $J_4$, while the dashed lines
representing the fits using only $J_1$. Inspection of
Fig.~\ref{fig:lcmo30dispersion} reveals that, as in LCMO25, the
inclusion of the $J_4$ gives a better fit to the data. The
normalized chi-squared, $\chi^2$, including $J_4$ is a factor of 2
less than that without it. The renormalization of spin wave
dispersions at large wavevectors is better seen in the $[\xi,0,0]$
and $[\xi,\xi,0]$ directions, where the boundary energies are
lowered to 25 and 52~meV, respectively. They are nearly 9 and 15~meV
lower than the expectation from a simple NN Heisenberg
Hamiltonian. We note that fits at larger energies ($E>70$~meV) are
not satisfactory, possibly because uncertainties associated with the
determination of a already damped magnetic excitation at high
energies.

\begin{table*}[ht!]
\begin{tabular}{llllccr}
\hline
\hline
Direction & $J_1$(meV) & $J_4$(meV)& $J_4/J_1$ & $D$($\rm meV \AA^2)$ & No.~of data sets &$\chi^2$\\
\hline
$[\xi,0,0]$    &6.01(5) &1.29(4) & 0.215(8) & $166.6\pm2.8$ & 13 & 1.71\\
$[\xi,\xi,0]$      &6.52(5) &1.18(4) &0.181(6) & $167.4\pm2.5$ & 22 & 1.91 \\
$[\xi,\xi,\xi]$    &6.56(9)&1.18(6) &0.179(11) & $167.8\pm4.8$ & 15 & 1.54 \\
$[\xi,0,0]+[\xi,\xi,0]+[\xi,\xi,\xi]$&6.36(3)&1.24(2) &0.195(4) & $168.6\pm1.7$ & 50 & 1.74 \\
$[\xi,0,0]+[\xi,\xi,0]+[\xi,\xi,\xi]$&8.43(3)&0(Fixed)&0   & $125.7\pm0.5$ & 50 & 3.25 \\
\hline
\hline
\end{tabular}
\caption{\label{tab:lcmo30}
Fitting parameters of exchange coupling constant $J_1$, $J_4$, the
ratio of $J_4/J_1$ and spin wave stiffness constant $D$ in LCMO30.
See Table~\ref{tab:lcmo25} for additional information.
}
\end{table*}

\begin{figure}[ht!]
\includegraphics[width=3.2in]{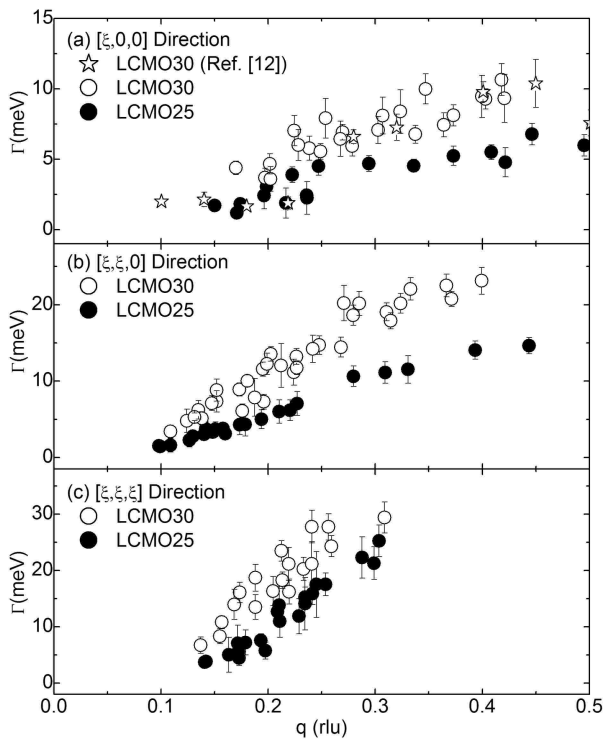}
\caption{\label{fig:gamma}
The wavevector dependence of intrinsic excitation widths
$\Gamma({\bf{q}})$ along three high symmetry directions in LCMO25
and LCMO30. The $q$-dependence of $\Gamma({\bf{q}})$ along
$[\xi,0,0]$ for LCMO30 obtained from triple-axis measurement
(Ref.~[\onlinecite{dai00phonon}]) is displayed for comparison.
}
\end{figure}

Having determined the dispersion relations for LCMO25 and LCMO30
using the model described in Eqn.\ \ref{eqn:dispersion} with
inclusion of $J_1$ and $J_4$, we now consider the other important
aspect of the spin dynamics, the intrinsic width
$\Gamma({\bf{q}})$. This linewidth $\Gamma({\bf{q}})$ is calculated
from Eqn.~4, where $\gamma$ is obtained from the best fit to
Eqn.~\ref{eqn:chi}. It is directly associated with the relevant
damping mechanisms and reflects how the quantized magnons interact
with other scattering processes. The wavevector dependence of
$\Gamma({\bf{q}})$ along different directions is shown in
Fig.~\ref{fig:gamma}. Note that there is marked difference between
LCMO25 and LCMO30; the widths of the latter are always larger,
indicating more damped excitations in LCMO30. In addition, the
momentum evolution of $\Gamma({\bf{q}})$ does not show any anomaly
across the Brillouin zone, appear to be isotopic and increase
drastically near the ZB. For example, $\Gamma({\bf{q}})$ reaches
around 20~meV for $E$ greater than 60~meV along the $[\xi,\xi,\xi]$
direction. This is unexpected from a classic, cubic Heisenberg
ferromagnet, where the magnetic excitations near the ZB are well
resolved.\cite{jaimeunpublished} The intrinsic widths in both LCMO25
and LCMO30 are much broader, indicating that one or more decay
mechanisms play an important role in these CMR compounds. 

We are now in a position to compare our data to various possible
mechanisms of the ZB softening. First, we want to comment on the
role of disorder or randomness which is naturally present in these
compounds. Motome and Furukawa \cite{motome02} have pointed out that
disorder in CMR materials will cause anomalous broadening and/or
anticrossing in the spin excitation spectra. In this scenario, the
one-electron bandwidth is proportional to the average ionic-size at
La/Ca site ($\bar{r}=\sum_{i}x_ir$, where $x_i$ is the fractional
occupancies of $A$-site species, $r_i$ is the individual radius). In
a previous paper, we have characterized the softening of dispersion
relation in a series of doped manganites near $x=0.30$
along $[\xi,0,0]$ direction.\cite{ye06} We found the renormalization
near ZB has little dependence on the average ionic size at $A$-site.
For LCMO25 and LCMO30, $\bar{r}$ become $1.207$ and $1.205$~\AA,
respectively, showing very little variation of oxygen octahedron
distortion surrounding the Mn-ions. On the other hand, the mismatch
between La and Ca ions will cause a quenched disorder in the system,
which can be qualitatively characterized by
$\sigma^2=\sum_{i}(x_ir^2_i-\bar{r}^2)$.\cite{martinez96,ionsize}
Base on this, $\sigma^2$ is $2.43\times 10^{-4}$ for LCMO25 and
$2.72\times 10^{-4}$ for LCMO30. Clearly, the differences in
quenched disorder between LCMO25 and LCMO30 are rather small and
cannot account for the dramatic change of spin wave spectra near the
ZB. We therefore conclude that the disorder effect does not play an
important role in this doping range of LCMO.

Second, the effect of magnon-phonon coupling may be the microscopic
origin of the observed magnon broadening and damping. Dai and
co-workers suggest that the interaction between optical phonon and
spin wave branches may lead to the broadening of spin wave spectrum
in a number of doped manganites near $x=0.30$.\cite{dai00phonon} In
this picture, a dispersionless optical phonon branch with energy
around 20~meV goes across the whole Brillouin zone and interacts
with the magnon branch, causing softening and a substantial increase
of the magnetic excitation linewidths. The measurements of Dai {\it
et al}.~were carried out with unpolarized neutrons, and it was
difficult to separate the magnetic scattering from the purely
lattice excitations. Recent measurements by Fernandez-Baca {\it
et~al.}\cite{jaime06} using polarized neutron scattering techniques
confirmed that the ZB spin waves of LCMO30 in the $[\xi,0,0]$ and
$[\xi,\xi,0]$ directions are considerable broad and have energies
lower than those expected from the approximation to the NN
Heisenberg Hamiltonian, although the softening in $[\xi,\xi,0]$ is
less as severe than originally reported.\cite{magneticform}  While
the magnon-phonon interaction  mechanism seems
to explain the broadening of the spin waves, it is not clear if it
would fully account for the magnitude of the observed softening in
LCMO30 at the zone boundary. Furthermore, this mechanism may not be
relevant to the general case of the manganites as the observed spin
waves in the $\rm Sm_{0.55}Sr_{0.45}MnO_3$ are softened to around
15~meV (which is below the 20~meV optical phonon) at the ZB along
$[\xi,0,0]$ direction, with no evidence of broadening.\cite{endoh05}

Finally, the distinct feature of ZB softening might be a consequence
of the $e_g$-band filling in the half-metallic
region.\cite{solovyev99} Solovyev and Terakura suggested that the
canonical double exchange is no longer appropriate as soon as holes
are doped into the system. Longer range FM interactions lead to the
softening at the zone boundary and contribute to the increase of the
stiffness constant $D$.\cite{solovyev99} However, the details of the
realistic electronic structure are important and may significantly
modify the analysis, particularly if the effects of $t_{2g}$
electrons are taken into account. For example, the $x$ dependence
of magnetic interactions is substantial modified by the change of DE
interaction contributed by $t_{2g}$ electrons and $D$ might even
decrease with $x$. Such complex scenario prevents us to make
meaningful comparison with the $D$'s yield experimentally. It is also
unclear whether this model can explain the commonly observed
damping/broadening of magnetic excitation near the ZB.  We would
like to point out that, in the tight-binding approximation used by these
authors, only the exchange couplings along the Mn-O-Mn chain
($J_1$, $J_4$, $J_8$, and $J_{15}$ as defined in
Ref.~\onlinecite{solovyev99}) would bring appreciable contributions
to the spin wave dynamics. Our work, and that of
Endoh,\cite{endoh05} show that $J_1$ and $J_4$ are the only exchange
constants contributing to the ZB softening. There is close
correlation between $J_i$'s and the orbital polarized
states;\cite{endoh05} $J_2$ would be enhanced by ($x^2-y^2$)-type
orbitals and $J_4$ is enhanced of ($3z^2-r^2$)-type orbitals. It is
surprising that a large increase of $J_4/J_1$ occurs despite of the
small change in nominal hole doping. This might indicate a drastic
modification of (preformed) orbital correlations which favor the
overlap between neighboring Mn ions.\cite{brink01} We hope that the
results presented here will help to stimulate further experimental
and theoretical investigations leading to a complete understanding
of the magnetic dynamics in those CMR materials. 

\section{Conclusions}

In summary, we have performed a systematic study on the spin wave
dynamics in the CMR manganites LCMO25 and LCMO30 using inelastic neutron
scattering techniques. We find that both systems display
considerable spin wave renormalization along all high symmetry
directions. Consistent with early measurements, which entirely focus
on the [1,0,0] direction, we find that the dispersion relations can
be phenomenologically analyzed using the NN interaction $J_1$ and
the $\rm 4^{th}$ NN interaction $J_4$. The introduction of $J_4$ 
lowers $J_1$ and therefore lowers the ZB energy. As a result of this
the systems with a larger $J_4/J_1$ ratio exhibit a larger SW
softening at the ZB.
The possible mechanisms responsible for such boundary softening are
also discussed. 

\section{Acknowledgments}

We thank Thomas Enck for preliminary data analysis. The authors are
grateful for the valuable discussion with Dr. M.~E. Hagen. This
work was supported by U.S. NSF DMR-0453804. ORNL is supported by
U.S. DOE under Contract No. DE-AC05-00OR22725 with UT/Battelle LLC.


\begin{references}
\bibitem{fitdetail} In this approach, only $J_1$ is allowed to vary
(the amplitude $A$ is fixed as its value has already been reliably
determined from the global fitting, $J_4$ is fixed to be zero), the
analysis would not carry the collective information of the whole
dispersion relation but does provide the correct energy at a
specific wavevector $q$.
\end{references}
\end{document}